\documentclass{iaus}
\usepackage{graphicx}
\usepackage{graphics}

\title[Triggered Star Formation in OB Associations] 
{Triggered Star Formation in OB Associations}

\author[Chen et al.]   
{W. P. Chen$^{1,2}$
 H. T. Lee$^1$ \and K. Sanchawala$^1$}

\affiliation{$^1$Institute of Astronomy, National Central 
University, Chung-Li 32054, Taiwan
\break email: wchen@astro.ncu.edu.tw, eridan@astro.ncu.edu.tw, 
kaushar@outflows.astro.ncu.edu.tw\\[\affilskip]
$^2$Department of Physics, National Central University, 
Chung-Li 32054, Taiwan}

\pubyear{2007}
\volume{237}  
\pagerange{}
\date{?? and in revised form ??}
\jname{Proceedings Title IAU Symposium}
\editors{A.C. Editor, B.D. Editor \& C.E. Editor, eds.}
\begin{document}

\maketitle

\begin{abstract}
We present causal and positional evidence of triggered star formation in 
bright-rimmed clouds in OB associations, e.g., Ori\,OB1, and Lac\,OB1, by 
photoionization.  The triggering process is seen also on a much larger scale in the 
Orion-Monoceros Complex by the Orion-Eridanus Superbubble.  We also 
show how the positioning of young stellar groups surrounding the H~II region
associated with Trumpler~16 in Carina Nebula supports the triggering process 
of star formation by the collect-and-collapse scenario.
\end{abstract}

\firstsection 
\section{Introduction}

A massive star has a profound influence on nearby molecular clouds.
On the one hand, the stellar radiation and energetic wind could evaporate 
the clouds and henceforth terminate the star-forming processes.  On the other 
hand, the 
massive star may provide ``just the touch'' to prompt the collapse of a molecular cloud 
which otherwise may not contract and fragment spontaneously.  Except perhaps in an 
environment such as the nucleus of a starburst galaxy, for which triggering predominates 
the star formation process, triggering in most cases likely plays a constructive albeit  
auxiliary role.  Other than providing additional push for cloud collapse, triggered 
star formation is self-sustaining (in time) and self-propagating (in space) in comparison 
to spontaneous cloud collapse.     

The extent to which a massive star influences the starbirth in a cloud depends on the 
amount and proximity of the cloud material, and the positional configuration 
between the massive star and the cloud.  If the massive star is born deep inside a  
cloud, which is often the case for a giant molecular cloud, the ionization 
fronts from the H~II region created by the massive star push the cloud from within, 
forming a cavity.  The gas and dust hence accumulate to a layer until the critical 
density is reached for gravitational collapse to form the next generation of stars.  
This is the so-called "collect-and-collapse" mechanism first proposed by 
\cite{elm77}, and recently demonstrated observationally by \cite{deh05} and \cite{zav06}.  
Any massive stars thus formed may subsequently break out their own cavities.  
Once a cavity forms, ionization now takes place on the surface of a remnant 
cloud.  Alternatively, the massive star and neighboring clouds could initially be 
already oriented in this way.  The UV photons from the massive star hence ionize 
the surface layer of the cloud, which illuminates as a bright rim seen prominently 
in an H-alpha image.  The ionization fronts embracing the surface of the cloud 
then result in a shock compressing into the cloud to cause the dense clumps to 
collapse.  This so-called "radiaton-driven implosion" (RDI) process has 
been proposed to account for triggered star formation in bright-rimmed clouds near 
H~II regions (\cite[Bertoldi 1989]{ber89}, \cite[Bertoldi \& McKee 1990]{ber90}, 
\cite[Hester \& Desch 2005]{hes05}).  A massive star at the end of its life, 
with its Wolf-Rayet winds and supernova explosion, may create a superbubble which 
can have an impact on even larger scales, tens or perhaps hundreds of parsecs away.  
Here we report on observations to illustrate triggered star formation by the 
collect-collapse-clear process and by the RDI mechanism in some OB associations. 

\section{Triggered Star Formation in Carina Nebula}

The Carina Nebula is known to contain the largest number of early-type
stars in the Milky Way, with a total of 64 O-type stars (\cite[Feinstein 1995]{fei95}).
Among the dozen known star clusters in the region, Trumpler\,14 and 
Trumpler\,16 are centrally located 
and are the youngest and the most populous.  These two clusters host 6 exceedingly 
rare main-sequence O3 stars.  In particular Trumpler\,16 contains a Wolf-Rayet star, 
HD\,93162, and the famous luminous blue variable, $\eta$ Carinae, which is arguably the most
massive star in our Galaxy (\cite[Massey \& Johnson 1993]{mas93}).  The Carina Nebula 
therefore serves as a unique laboratory to study not only the massive star formation process, but 
also the interplay among massive stars, interstellar media and low-mass star formation.

\begin{figure}
 \centering
 \includegraphics[height=3in,width=3in]{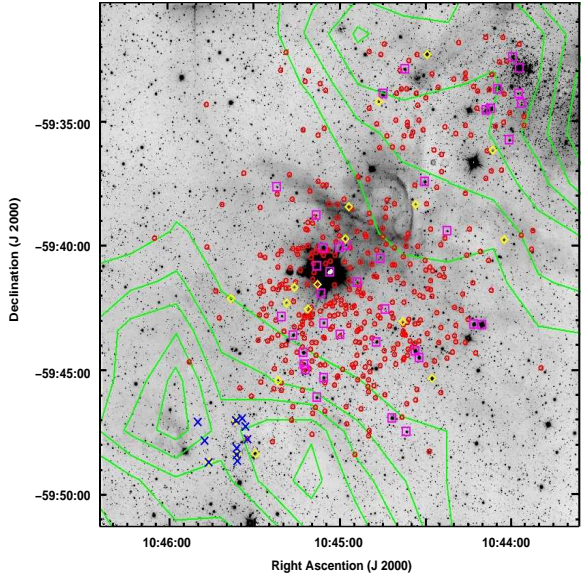}
  \caption{
    The $Chandra$ sources overlaid on the mosaic $K_s$ band image 
    (\cite[Sanchawala et al. 2006]{san06}) taken by 
the IRSF, centered on $\eta$ Carinae, with the contours showing the
$^{12}CO(1-0)$ emission (\cite[Brooks et al. 1998]{bro98}). Known OB stars are marked as
boxes and the candidate OB stars are marked as diamonds.
The stars of the embedded X-ray group to the south-east of Trumpler\,16 are
marked as crosses, whereas all other $Chandra$ sources 
are marked as circles.
          }
  \label{fig:carina}
\end{figure}

\cite{san06} studied the X-ray sources detected by $Chandra$ in the Carina Nebula 
(Fig.~\ref{fig:carina}).  
Of the 454 X-ray sources, 38 coincide with known OB stars.  Additionally, 16  
anonymous stars have been found to have X-ray and near-infrared properties similar 
to those of the known OB stars.  These candidate OB stars likely escaped 
earlier optical studies because of their excessive dust extinction.  
Close to 200 X-ray sources are candidate 
classical T Tauri stars (CTTSs), judged on the basis of their infrared colors.  This 
sample represents the most comprehensive census of the young stellar population 
in the Carina Nebula so far and is useful for the study of the 
star-formation history in this turbulent environment.  

In Fig.~\ref{fig:carina}, in addition to Trumpler\,16 near the center 
and Trumpler\,14 
to the north-west, there is an embedded group of 10 young stars 
to the south-east of Trumpler\,16, 'sandwiched' between two cloud 
peaks (\cite[Sanchawala et al. 2006]{san06}).  One sees immediately the
general paucity of massive stars with respect to molecular clouds; namely   
Trumpler\,16 itself has cast out a cavity and lies between the north-west 
and south-east cloud complexes, and so has Trumpler\,14 to a less extent.  The newly 
identified group suffers a large amount of reddening and also is situated 
between cloud peaks, apparently in the initial stage to expel the gas.  
There seems a general tendency for the X-ray sources 
(i.e., young stars) to be either intervening between clouds or located near the cloud 
surfaces facing Trumpler\,16.   The morphology of young stellar groups and molecular 
clouds peripheral to an H~II region (i.e., Trumpler\,16 here) fits closely the 
description of the collect-and-collapse mechanism for massive star formation.  

\section{Triggered Star Formation in Orion OB1, near $\lambda$ Ori, and Lac OB1}

\begin{figure}
  \centering
  \includegraphics[height=2in,width=5in]{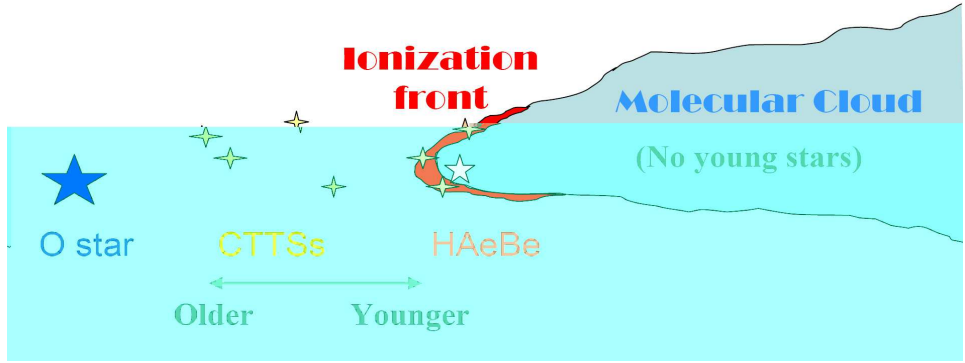}
  \caption{An illustration of a massive star to trigger star formation in
           a nearby molecular cloud.}
  \label{fig:starcloud}
\end{figure}

An RDI triggering process would leave several imprints that can be diagnosed 
observationally (Fig.~\ref{fig:starcloud}):
(1)~The remnant cloud is extended toward, or pointing to, the massive
stars.  (2)~The young stellar groupings are roughly lined up between the
remnant clouds and the luminous star, (3)~Stars closer to the cloud,
formed later in the sequence, are younger in age, with the youngest
distributed at the interacting region, i.e., along the bright rim of a cloud,
and (4)~No young stars exist far inside the cloud, i.e., leading the ionization 
front.   In particular, the temporal and positional signposts, (3) and (4), 
are in distinct contrast to the case of spontaneous star formation by a global cloud 
collapse, which would lead to starbirth spreading throughout the cloud.  

\begin{figure}
  \centering
  \includegraphics[height=3.2in,width=2.4in,angle=-90]{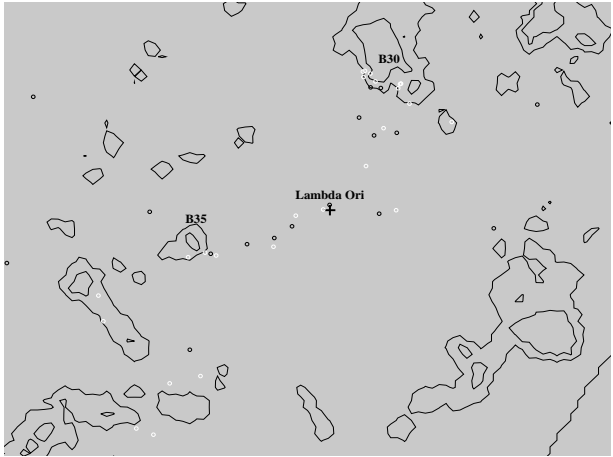}
  \caption{The CTTSs appear to line up between $\lambda$\,Ori and remnant clouds, with 
            the clouds roughly pointing to $\lambda$\,Ori 
            (\cite[Lee et al. 2005]{lee05}). }
  \label{fig:lambdaco}
\end{figure}

\cite{lee05} and \cite{lee06a} developed an empirical set of criteria, 
based on the Two Micron All Sky Survey (2MASS) colors, to select CTTSs and 
Herbig Ae/Be (HAeBe) stars in star-forming regions.  The selection 
criteria prove 
very effective as follow-up spectroscopy showed that most candidates 
associated with nebulosity were indeed young stars, but otherwise were carbon 
stars or M giants.  With the sensitivity of 2MASS, young stars within
$\sim 1$~kpc can be readily recognized.  In Ori\,OB1, $\lambda$ Ori, and Lac\,OB1 
there is compelling evidence of RDI triggering to form low- and intermediate-mass 
stars.  Fig.~\ref{fig:lambdaco} shows an example near $\lambda$\,Ori.  Analysis of 
2MASS colors shows the young stars to be progressively younger toward the clouds, 
with the youngest near the cloud rim.  Furthermore, there is a tendency for 
HAeBe stars to reside deeper into the cloud, indicating that more massive stars, 
when prompted to form, appear to favor denser environments where photoevaporation 
effect is reduced.  On the other hand, when a dense core near the ionization 
layer (i.e., current cloud surface) collapses, the accretion process has to 
compete with the mass loss arising from photoevaporation, leading to formation of 
a less massive star or even substellar objects (\cite[Whitworth \& 
Zinnecker 2004]{whi04}).  Eventually the 
remnant cloud would be dispersed completely, and stars of different masses remain in 
the same volume.  On a larger scale, the Wolf-Rayet winds and supernova explosion of 
a massive star would create a superbubble ramping on one molecular cloud to another 
(\cite[Lee \& Chen 2006b]{lee06b}).  
A sequence of such events ("relay star formation") could spread the star formation 
out to tens or even hundreds of parsec away.  Note that the CTTS sample traces only recent star 
formation, and the bright-rimmed clouds present convenient snapshots to show how triggering 
leads to formation of low- and intermediate-mass stars once an O star is formed.  
These processes do not preclude stars formed in earlier epochs, with whatever 
mechanisms and masses, which already exist in a region.         

\begin{acknowledgments}
We acknowledge the grant NSC94-2112-M-008-017 from the National Science Council of Taiwan to 
support this research. 
\end{acknowledgments}


\begin{thebibliography}{}
%

\bibitem[Bertoldi (1989)]{ber89} 
          {Bertoldi, F.} 1989, \textit{ApJ}, 346, 735
\bibitem[Bertoldi \& McKee(1990)]{ber90} 
          {Bertoldi, F. \& McKee, C. F.} 1990, \textit{ApJ}, 354, 529
\bibitem[Brooks et al. (1998)]{bro98} 
          {Brooks, K., Whiteoak, J. B., \&
           Storey, J. W. V.} 1998, \textit{PASA}, 15, 202
\bibitem[Deharveng, Zavagno \& Caplan (2005)]{deh05} 
          {Deharveng, L., Zavagno, A., \& Caplan, J.} 2005, \textit{A \& A}, 433, 565
\bibitem[Elmegreen \& Lada (1977)]{elm77} 
          {Elmegreen, B. G. \& Lada, C. J.} 1977, \textit{ApJ}, 214, 725
\bibitem[Feinstein (1995)]{fei95} Feinstein, A. 1995, \textit{RevMexAA}, 2, 57
\bibitem[Hester \& Desch(2005)]{hes05} 
          {Hester, J.~J., \& Desch, S.~J.} 2005, ASP Conf.~Ser.~341:~Chondrites and the 
             Protoplanetary Disk, 341, 107
\bibitem[Lee et al. (2005)]{lee05} 
          {Lee, H.-T., Chen, W. P., Zhang, Z. W., \& Hu, J. Y.} 2005, \textit{ApJ}, 624, 808
\bibitem[Lee \& Chen (2006a)]{lee06a}
          {Lee, H. T., \& Chen, W. P.} 2006, \textit{ApJ in submission}, astro-ph/0509315
\bibitem[Lee \& Chen (2006b)]{lee06b}
          {Lee, H. T., \& Chen, W. P.} 2006, \textit{AJ in submission}, astro-ph/0608216
\bibitem[Massey \& Johnson (1993)]{mas93} 
          {Massey, P., \& Johnson, J.} 1993, \textit{AJ}, 105, 980
\bibitem[Sanchawala et al. (2006)]{san06} 
          {Sanchawala, K. et al.} 2006, \textit{ApJ in submission}, astro-ph/0603043
\bibitem[Whitworth \& Zinnecker (2004)]{whi04} Whitworth, A.~P., \& Zinnecker, H.
          2004, \textit{A \& A}, 427, 299
\bibitem[Zavagno et al. (2006)]{zav06} 
          {Zavagno, A., Deharveng, L., Comer\'{o}n, F., Brand, J.,
               Massi, F., Caplan, J., \& Russeil, D.} 2006 \textit{A \& A}, 446, 171
%
\end{thebibliography}
\end{document}